\documentstyle[amssymb,multicol,epsfig,aps,prl]{revtex}

\begin{document}
\title{ ELECTRONIC STRUCTURE, ELECTRON-PHONON COUPLING, AND MULTIBAND
EFFECTS IN MgB$_{2}$}
\author{I.I. Mazin$^\dagger$ and V.P. Antropov$^*$}
\address{$^\dagger$Code 6391, Naval Research Laboratory, Washington, DC 
20375\\
$^*$Ames Laboratory, Ames, IA, 50011\\
}
\date{10/11/2002}

\maketitle
\begin{abstract}
We review the current situation in the theory of superconducting 
and transport properties of MgB$_2$. First principle calculations of 
of the electronic structure and electron-phonon coupling are
discussed and compared with the experiment. We also
present a brief description of the multiband effects in superconductivity
and transport, and how these manifest themselves in MgB$_2$.
We also mention some yet open questions. 
\end{abstract}

\begin{flushright}
{\it Is there anything of which one can say: 
 ``Look! This is something new''?  \\
It was here already, long ago;  
it was here before our time. \\
Ecclesiastes, 1:10. }
\end{flushright}
\begin{multicols}{2}
\section{Introduction}

Many of us remember that fabulous excitement that reigned in physics world
after the discovery of high-$T_c$ cuprates. Since then, we have become so
familiar with record-breaking temperatures of 90 K, 120 K, 160 K, that it is
worth recalling that 15 years ago not only the highest known superconducting
temperature was meager 24 K, but it was also believed by many since early
70's \cite{CA} that this temperature is close to the theoretical limit
for electron-phonon superconductivity.

High-$T_c$ superconductivity revolutionized our approaches both to theory
and to experiment. However, in the shadow of mysterious cuprates
lower-temperature superconductors were receiving relatively little attention.

This has been changed recently. In 2001 alone, besides the report of 40 K
superconductivity in the simple magnesium diboride, exciting cases of
superconductivity coexisting with magnetism (ZrZn$_2$), possibly induced by
magnetism ($\varepsilon$-Fe), or competing with magnetism (MgCNi$_3$) were
reported. While all these cases are different and probably manifest quite
different physics, all of them indicate that the physics community turned
its face back to low-temperature superconductivity. And, of course, MgB$_2$
is the champion of the year, hands down.

Very similar to the high-$T_c$ cuprates, immediately after its discovery\cite{A}
some authors described MgB$_2$ as an extreme case of conventional,
``Eliashberg'' superconductivity, an extremely lucky combination of the
fortunate parameters\cite{us,An}, while the others suggested variety of
exotic electronic mechanisms, possibly similar to cuprates\cite%
{hirsch,rice,baskaran,yamaji,marsiglio}. But the analogy stops here. Now,
two years after the discovery, we already have much better understanding and
much more universal consensus about the physics of MgB$_2$, than about
cuprates. In fact, an agreement emerges that it is, albeit still an
electron-phonon superconductor, a case of genuinely novel physics,
sufficiently unusual to set it apart from all previous electron-phonon
superconductors\cite{Liu}.

One of the main factors that distinguishes  MgB$_2$  from the high-$T_c$ cuprates
is that the electronic structure of this materials is very well described by
conventional band-theoretical methods, which have been perfected in the last
decades to the level that allows unprecedentally detailed first-principle
calculations of electron and phonon spectra, and of the electron-phonon
calculations. Excellent agreement of such {\it ab initio} calculations with
the experiment literally leaves hardly any room to play with exotic, but
hardly verifiable models, so popular in the high-$T_c$ world. In this
Chapter we will try to present a broad view on the physics of MgB$_2$, as it
currently emerges from the first-principle calculation, and seems to be
fully supported by the experiment.

The Chapter is organized as follows: The electronic structure of bulk MgB$%
_{2} $ is disussed in Section 2, which also deals with some experiments that
give credit to the calculated band structure. In Section 3 we discuss first
principles calculations of the phonon spectra and the electron-phonon
coupling (EPC). Section 4 is devoted to the discussion of multiband effects
in MgB$_{2}.$

\section{Electronic structure}

\subsection{General description}

MgB$_{2}$ occurs in the AlB$_{2}$ structure. Boron atoms reside in
graphite-like (honeycomb) layers stacked with no displacement~\cite{MGBSTR}
forming hexagonal prisms with the base translation almost equal to the
height, $a=3.085$ (3.009) \AA\ and $c/a=1.142$ (1.084) for MgB$_{2}$ (AlB$%
_{2}$). These prisms contain large, nearly spherical pores occupied by Mg
atoms. This structure may therefore be regarded as that of completely
intercalated graphite~\cite{Bur} with carbon replaced by boron, its neighbor
in the periodic table. Furthermore, MgB$_{2}$ is formally isoelectronic to
graphite. Therefore, chemical bonding and electronic properties of MgB$_{2}$
are expected to have some similarity to those of graphite and graphite
intercalation compounds, some of which also exhibit superconductivity. As in
graphite ($R_{{\rm intra}}$=1.42~\AA ), the intralayer B--B bonds are much
shorter than the interlayer distance, and hence the B--B bonding is strongly
anisotropic. However, the intralayer bonds are only twice as short as the
interlayer ones, compared to the ratio of 2.4 in graphite, allowing for a
significant interlayer hopping. For comparison, the interatomic distance
between nearest neighbors is 1.55~\AA\ in diamond and 1.4--1.45~\AA\ in the C%
$_{60}$ molecule.

In spite of a structural similarity to intercalated graphite and, to some
extent, to doped fullerenes, MgB$_2$ has a qualitatively different and
rather uncommon structure of the conducting states setting it aside from
both these groups of superconductors. The peculiar and (so far) unique
feature of MgB$_2$ is the incomplete filling of the two $\sigma$ bands
corresponding to strongly covalent, $sp^2$-hybrid bonding within the
graphite-like boron layer. The holes at the top of these $\sigma$ bands
manifest notably two-dimensional properties and are localized within the
boron sheets, in contrast with mostly three-dimensional electrons and holes
in the $\pi$ bands, which are delocalized over the whole crystal. These 2D
covalent and 3D metallic-type states contribute almost equally to the total
density of states (DOS) at the Fermi level, while the unfilled covalent
bands experience strong interaction with longitudinal vibrations in the
boron layer.

The band structure of MgB$_{2}$ had been reported long before the discovery
of superconductivity \cite{Tupitsyn,ArmstrongMgB2,Freeman-review,Medv} and
is now known in very detail. The results discussed in this Chapter were
obtained using LMTO-ASA, full-potential LMTO, or full-potential LAPW method.
Computational details may be found in respective original publications. For
MgB$_{2}$, there is usually little difference between different methods, in
any event, none important for the qualitative discussions in this Chapter.

\begin{figure}[tbp]
\centerline{
\epsfig{file=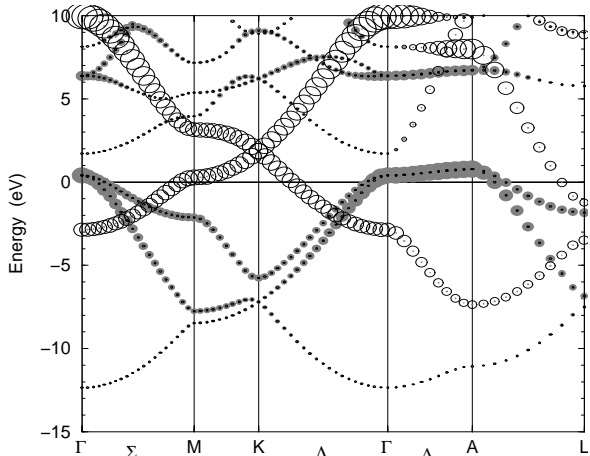,width=0.9\linewidth,clip=true}}
\setlength{\columnwidth}{\linewidth} \nopagebreak
\caption{Bandstructure of MgB$_2$ with the B p-character. The radii of the
hollow (filled) circles are proportional to the $\protect\pi$ ( $\protect%
\sigma$) character.}
\label{mgb2bnds}
\end{figure}
\begin{figure}[tbp]
\centerline{
\epsfig{file=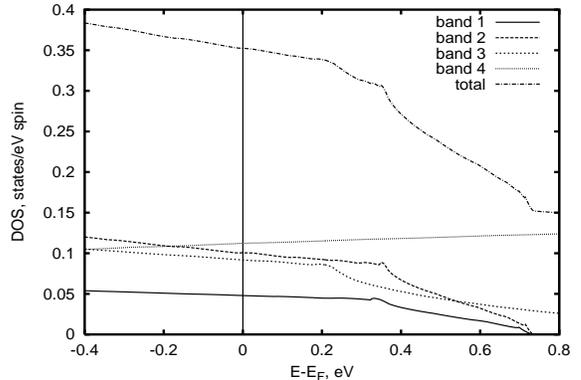,width=0.9\linewidth,height=0.6\linewidth,clip=true}
}
\setlength{\columnwidth}{\linewidth} \nopagebreak
\caption{Total density of states (DOS) and partial DOS for MgB$_2$. Bands
1,2 are $\protect\sigma$ bands, bands 3,4 are $\protect\pi$ bands }
\label{DOS}
\end{figure}

\begin{figure}[tbp]
\centerline{
\epsfig{file=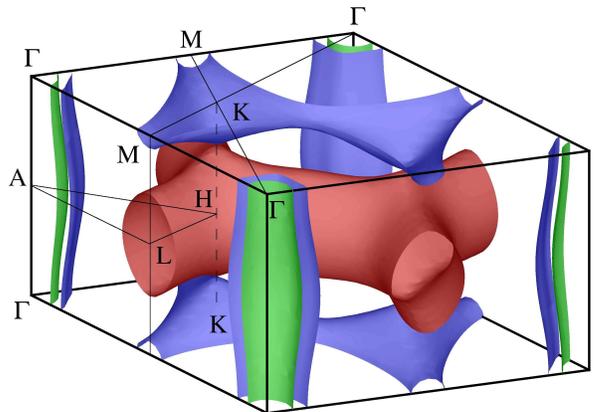,height=0.9\linewidth, angle=-90,
clip=true}
}
\setlength{\columnwidth}{\linewidth} \nopagebreak
\caption{Fermi surface of MgB$_2$.}
\label{FermiSurfaces}
\end{figure}

The energy bands, DOS and the Fermi surface of MgB$_{2}$ are shown in Figs.~%
\ref{mgb2bnds},~\ref{DOS} and~\ref{FermiSurfaces}. As expected, the bands
are quite similar to those of graphite with three bonding $\sigma $ bands
corresponding to in-plane $sp_{x}p_{y}$ ($sp^{2}$) hybridization in the
boron layer and two $\pi $ bands (bonding and antibonding) formed by
aromatically hybridized boron $p_{z}$ orbitals. Both $\sigma$ and $\pi$
bands have strong in-plane dispersion due to the large overlap between all $%
p $ orbitals (both in-plane and out-of-plane) for neighboring boron atoms.
The interlayer overlaps are much smaller, especially for $p_{xy}$ orbitals,
so that the $k_{z}$ dispersion of $\sigma $ bands does not exceed 1~eV. On
the other hand, in contrast to intercalated graphites, two of the $\sigma $
bands are filled incompletely. Together with weak $k_{z}$ dispersion this
results in the appearance of two nearly cylindrical sheets of the Fermi
surface (see Fig.~\ref{FermiSurfaces}) around the $\Gamma $--A line. As we
will see below from the analysis of the charge density distribution, these
unfilled $\sigma $ bands with boron $p_{xy}$ character fully retain their
covalent structure. Conducting covalent bonds represent a peculiar feature
of MgB$_{2}$ making it an exotic compound probably existing on the brink of
structural instability.

It is seen in Fig.~\ref{FermiSurfaces} that the $\pi$ bands form two planar
honeycomb tubular networks: an antibonding electron-type sheet centered at $%
k_{z}=0$ (red) and a similar, but more compact, bonding hole-type sheet
centered at $k_{z}=\pi /c$ (blue). These two sheets touch at some point on
the K-H line. The hole-type sheet is close to an electronic topological
transition (ETT) at the M point corresponding to the breakdown of the
tubular network into separate starfish-like pockets (at 0.25~eV above $E_{F}$%
).

\begin{figure}[tbp]
\centerline{
\epsfig{file=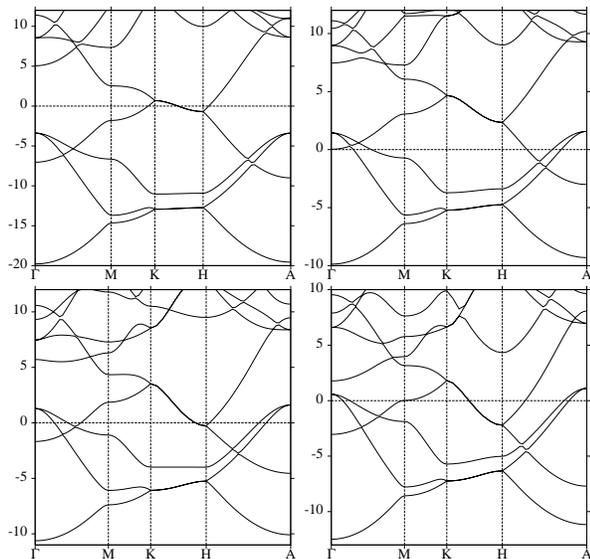,width=0.9\linewidth,angle=0,clip=true}
}
\setlength{\columnwidth}{\linewidth} \nopagebreak
\caption{Band structures of: (a) top left: primitive (AA stacking) graphite
(PG), $a=2.456$\AA , $c/a=1.363$; (b) top right: PG boron, $a=3.085$\AA , $%
c/a=1.142$ (as in MgB$_2$); (c) bottom left: LiB$_2$ in MgB$_2$ structure,
same $a$ and $c/a$; (d) bottom right: MgB$_{2}$, same $a$ and $c/a$. Energy
is in eV relative to $E_{F}$. The order of occupied bands in the $\Gamma $\
point is $\protect\sigma$ bonding with boron $s$ character, $\protect\pi$
bonding with boron $p_z$ character, and $\protect\sigma$ bonding with boron $%
p_{xy}$ character (double degenerate). }
\label{manybands}
\end{figure}

In order to examine the relation between the band structure of MgB$_{2}$\
and that of graphite in more detail one can compare the following
hypothetical sequence of intermediate materials: carbon in the `primitive
graphite' (PG) lattice with no displacement between layers as in MgB$_{2}$,
using graphite lattice parameters; boron in the PG lattice with $a$\ as in
MgB$_{2}$\ and $c/a$ as in graphite; boron in the PG lattice with $a$\ and $%
c/a$\ as in MgB$_{2}$; LiB$_{2}$ in the same structure; MgB$_{2}$\ itself.
The results of some of these calculations \cite{us2,REVMGB2} are shown in
Fig.~\ref{manybands}.

The band structure of PG carbon shown in Fig.~\ref{manybands}a is very
similar to that of graphite~\cite{Freeman} with the appropriate zone-folding
for a smaller unit cell. (This is quite natural because of the weak
interlayer interaction.) Boron in the same lattice (not shown) has nearly
identical bands with the energies scaled by the inverse square of the
lattice parameter, in agreement with canonical tight-binding scaling\cite%
{Harrison}. Fig.~\ref{manybands}b shows the natural enhancement of the
out-of-plane dispersion of the $\pi $ bands when the interlayer distance is
reduced. Figs.~\ref{manybands}c and \ref{manybands}d demonstrate that
`intercalation' of boron by Li or Mg produces a significant distortion of
the band structure, so that the role of the intercalant is not simply
donating electrons to boron's bands (which would recover the band structure
of PG carbon shown in Fig.~\ref{manybands}a). The main change upon
intercalation is the downward shift of the $\pi $ bands compared to $\sigma $
bands. For Li this shift of $\sim $1.5 eV is almost uniform throughout the
Brillouin zone. Replacement of Li by Mg shifts the $\pi $ bands further, but
this shift is strongly asymmetric increasing from $\sim $0.6 eV at the $%
\Gamma $ point to $\sim $ 2.6 eV at the A point. In addition, the
out-of-plane dispersion of the $\sigma $ bands is also significantly
enhanced. In LiB$_{2} $ the filling of the bonding $p_{xy}$ bands is nearly
the same as in PG boron, while in MgB$_{2}$ the Fermi level shifts closer to
the top of these bands.

The lowering of the $\pi$ bands in MgB$_{2}$ compared to PG boron is due to
stronger interaction of boron $p_{z}$ orbitals with ionized magnesium
sublattice compared to $p_{xy}$ orbitals. This lowering is greater at the
AHL plane compared to the $\Gamma $KM plane, because the antisymmetric (with 
$k_{z}=\pi /c$) overlap of the boron's $p_{z}$ tails increases the
electronic density close to the magnesium plane where its attractive
potential is the strongest. 
\begin{figure}[tbp]
\centerline{
\epsfig{file=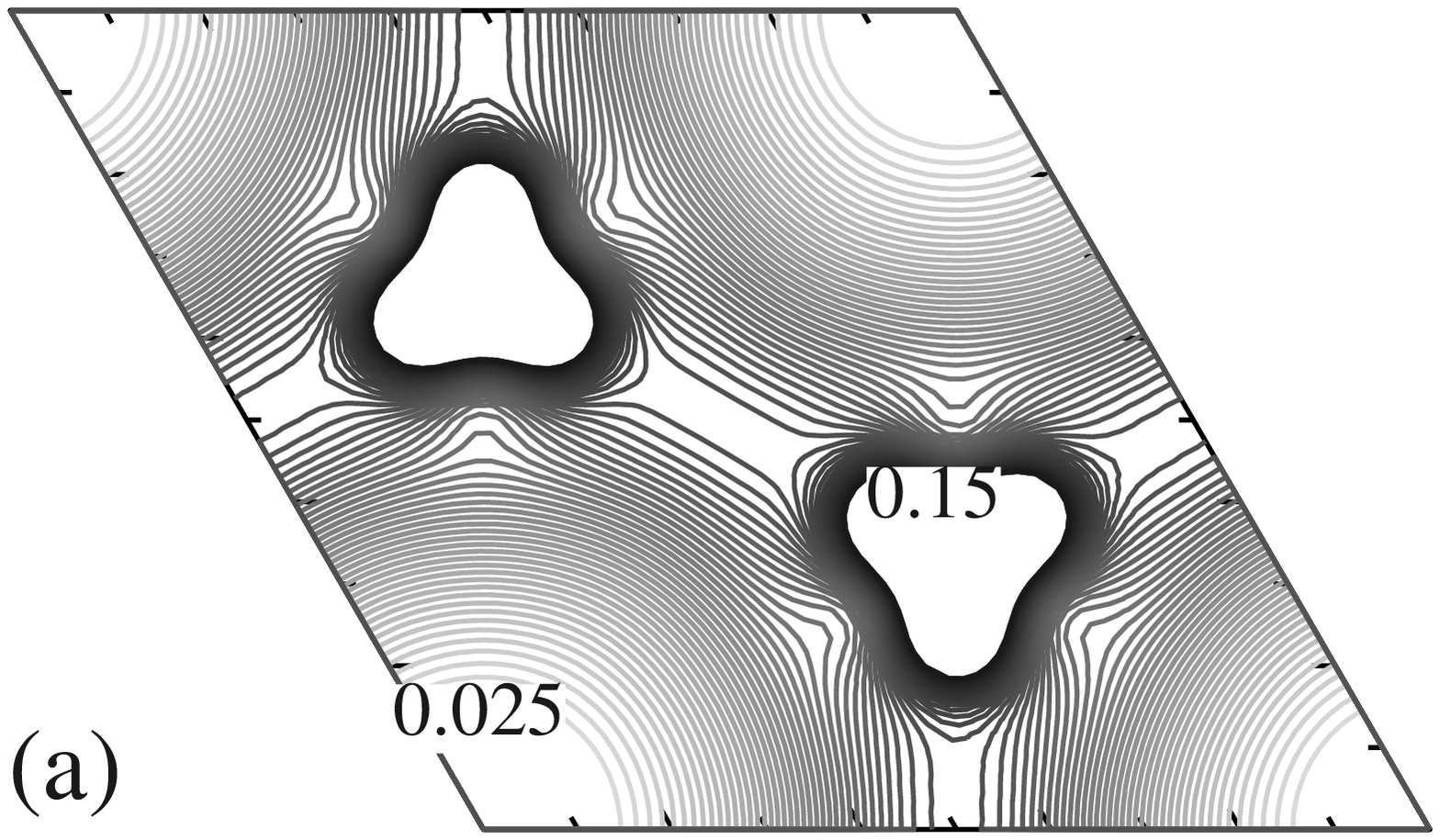,width=0.4\linewidth,angle=0,clip=true}
\epsfig{file=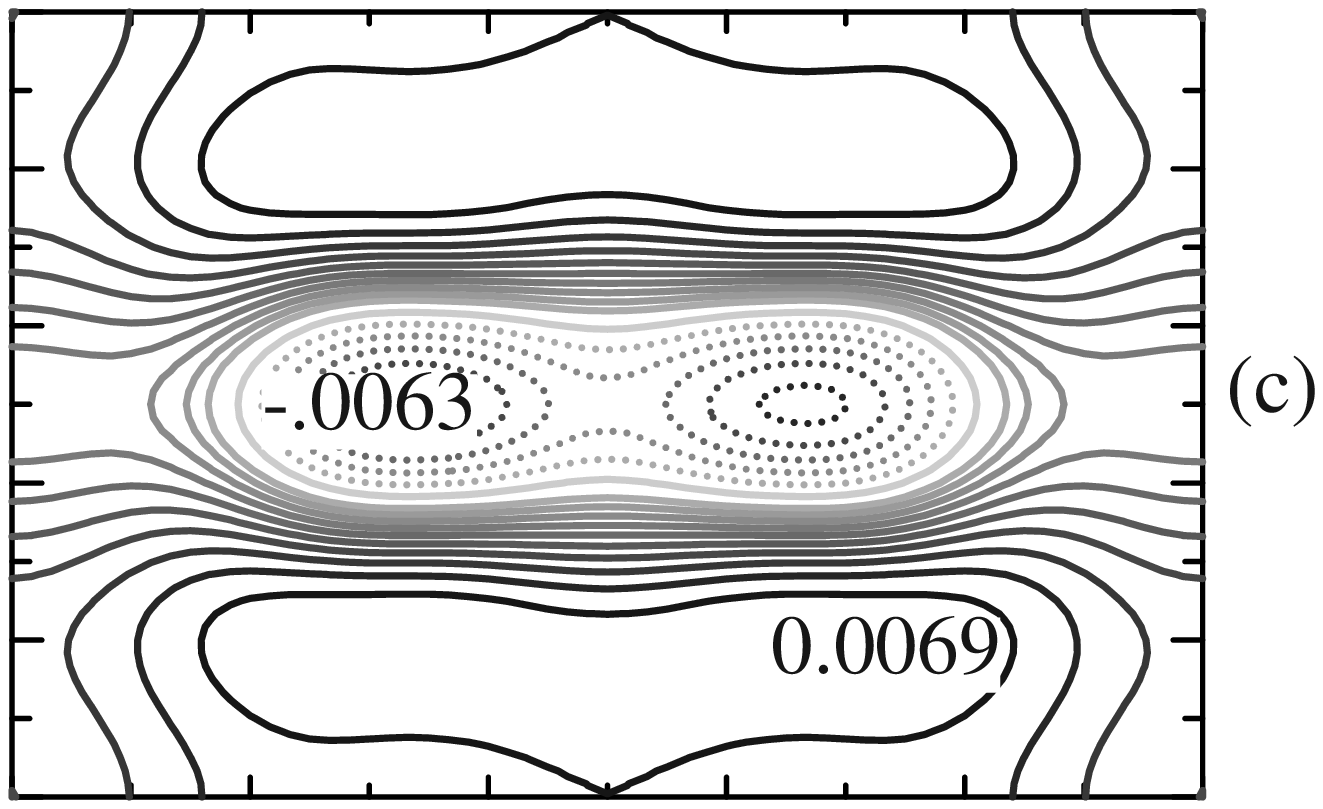,width=0.5\linewidth,angle=0,clip=true}}
\centerline{
\epsfig{file=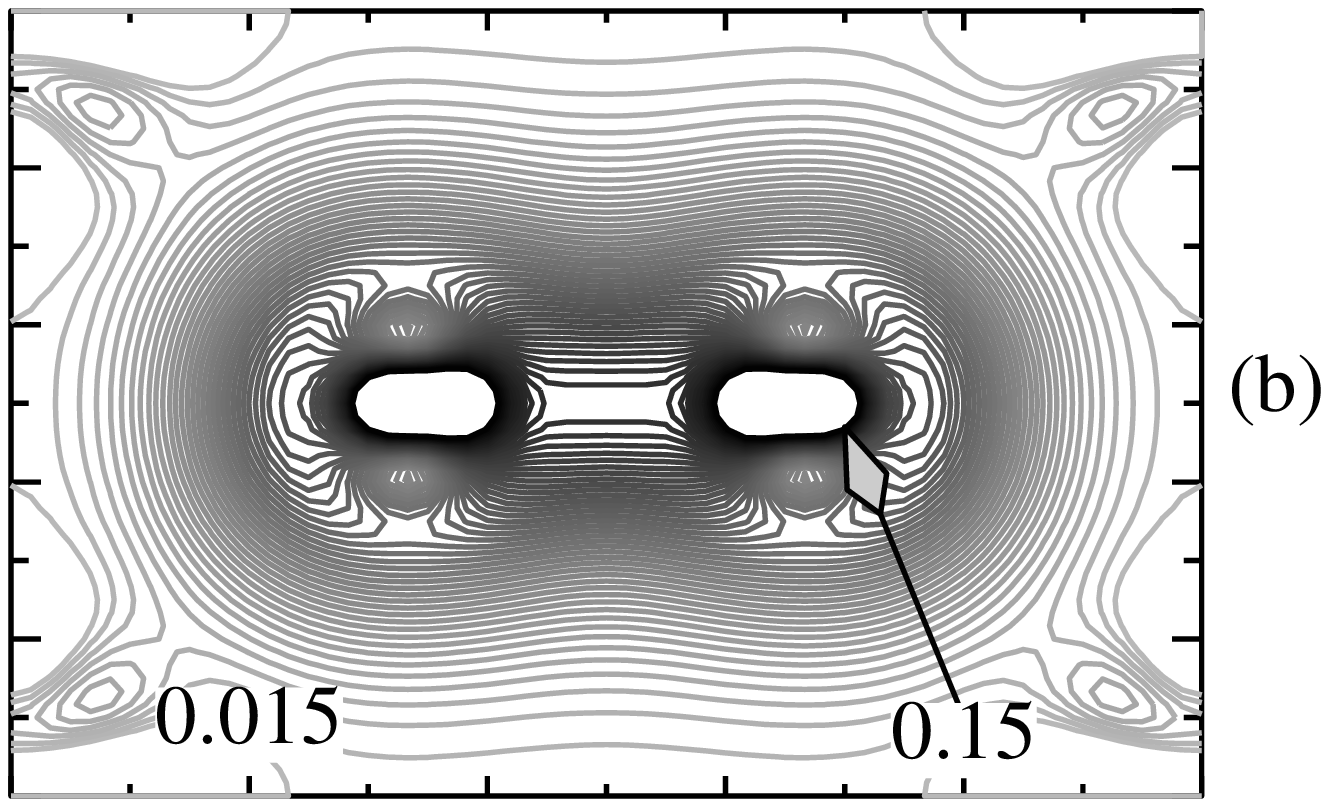,width=0.5\linewidth,angle=0,clip=true}
\epsfig{file=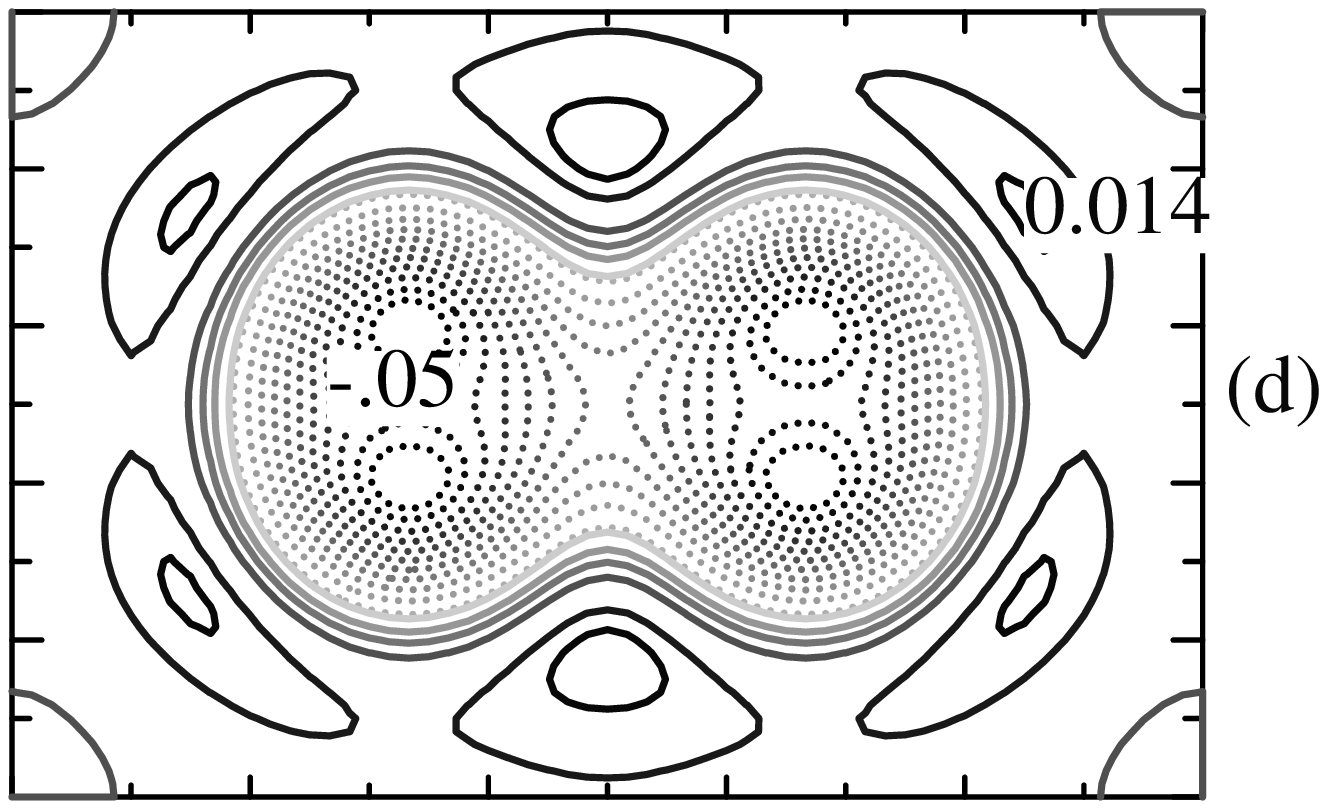,width=0.5\linewidth,angle=0,clip=true}}
\setlength{\columnwidth}{\linewidth} \nopagebreak
\caption{Pseudocharge density contours obtained in FLMTO. The unit cell is
everywhere that of MgB$_{2}$. Darkness of lines increases with density. (a)
MgB$_{2}$\ in (0002) plane passing through B nuclei; (b) MgB$_{2}$\ in
(1000) plane passing through Mg nuclei at each corner of the figure. B
nuclei occupy positions (1/3,1/2) and (2/3,1/2) in the plane of the figure.
The integrated charge of the unit cell is 8. (c) (1000) plane, difference in
smoothed density, MgB$_{2}$\ minus NaB$_{2}$. The integrated charge of the
unit cell is 1. (d) (1000) plane, difference in smoothed density, MgB$_{2}$\
minus PG carbon. The integrated charge of the unit cell is 0. In (c) and
(d), dotted lines show negative values. }
\label{charge}
\end{figure}

The nature of bonding in MgB$_{2}$ may be understood from the charge density
(CD) plots\cite{REVMGB2} shown in Fig.~\ref{charge}. As it is seen in Fig.~%
\ref{charge}a, bonding in the boron layer is typically covalent. The CD of
the boron atom is strongly aspherical, and the directional bonds with high
CD are clearly seen (see also Ref.~~\cite{Medv}). The CD distribution in the
boron layer is very similar to that in the carbon layer of graphite~\cite%
{Freeman}. This directional in-plane bonding is also obvious from Fig.~\ref%
{charge}b showing the CD in the cross section containing both Mg and B
atoms. However, Fig.~\ref{charge}b also shows that a large amount of valence
charge does not participate in any covalent bonding, but is rather
distributed more or less homogeneously over the whole crystal. Further, Fig.~%
\ref{charge}c shows the difference of the CD of MgB$_{2}$\ and that of
hypothetical NaB$_{2}$\ in exactly the same lattice. Not only does it show
that one extra valence electron is not absorbed by boron atoms but that it
is rather delocalized in the interstitials; it also shows that some charge
moves away from the boron atoms and covalent in-plane B-B bonds. Fig.~\ref%
{charge}d shows the CD difference between the isoelectronic compounds MgB$%
_{2}$\ and PG carbon (C$_{2}$). In MgB$_{2}$, the electrons see
approximately the same external potential as in C$_{2}$, except that one
proton is pulled from each C nucleus and put at the Mg site. It is evident
that the change C$_{2}${}$\rightarrow $MgB$_{2}$ weakens the two-center $%
\sigma $\ bonds (the charge between the atoms is depleted) and redistributes
it into a delocalized, metallic density.

\begin{figure}[tbp] \centerline{
\epsfig{file=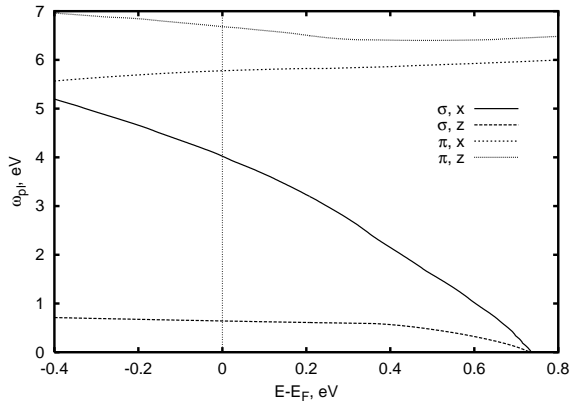,width=0.9\linewidth,angle=0,clip=true} }
\setlength{\columnwidth}{\linewidth} \nopagebreak
\caption{Plasma frequencies for $\protect\sigma $ and $\protect\pi$ bands.}
\label{vel} \end{figure}
A numerical reconstruction of the electronic charge density from the
synchrotron radiation data for a powder MgB$_2$ sample~\cite{CDexp} supports
this general picture. The charge density found for 15~K is, in fact, very
similar to that in Fig.~\ref{charge}b and shows all the important features
discussed above including the distinct covalent bonds within the boron
sheets, the strongly ionized Mg, and the delocalized charges in the
interstitials. Further, the Fourier maps obtained~\cite{MGBSTR} for the
single crystals also clearly show the covalent $sp^{2}$ hybrids in the boron
layer and no covalent bonding between B and Mg atoms.

Thus, one can say that MgB$_2$ is held together by strongly {\it covalent}
bonds within boron layers and by delocalized, `{\it metallic}-type' bonds
between these sheets. A peculiar feature of this compound is that electrons
participating in both of these bond types provide comparable contributions
to $N$. This distinguishes MgB$_2$ from closely related graphites where
covalent bonds in the carbon layers are always completely filled, while the
nearly cylindrical parts of the Fermi surface commonly found in those
compounds are formed by carbon-derived $\pi$ bands which are much less 3D
that the corresponding bands in MgB$_2$\cite{REVIEW}. 

Because of the coexistence of two different types of conducting states, one
needs to see the contributions to the total DOS and transport properties
from separate sheets of the Fermi surface originating from 2D covalent and
3D metallic-type bonding. This decomposition is shown in Fig.~\ref{DOS}
for the DOS and in Fig. \ref{vel} for the in-plane ($xx$) and out-of-plane ($%
zz$) components of the plasma frequency $\omega_{pl\
\alpha}^2=(e^2/2\pi^2)\int v^2_{\alpha} \delta[\epsilon ({\bf k})-E_F]d{\bf k%
}$, where $v_{\alpha}$ is the $\alpha$-component of the Fermi velocity. The
3D (metallic-type bonding) and cylindrical (covalent bonding) parts of the
Fermi surface contribute, respectively, about 58\% and 42\% to $N(E_F)$. If
the $\sigma$ Fermi surfaces were ideal cylinders, $N(E)$ for these bands
would have a step-like singularity at some 0.5 eV above $E_F$. This is
broadened by a nonzero $z$-dispersion. The hole $\pi$-band has a 3D van Hove
singularity in the same range of energies, while the electron-like $\pi$%
-band has a DOS which is rather flat around $E_F$. $\pi$-bands contribute
about 80\% to the total $\omega_{pl}^2$, and thus, given the same relaxation
rate for all bands, to total conductivity. While the total conductivity is
more or less isotropic, the $\sigma-$band conductivity is, as expected,
highly anisotropic.

\subsection{Experimental probes of electronic structure}

It is well known that in some materials conventional band structure
calculations do not reproduce the exterimental one-electron excitation
spectra with sufficient accuracy. These case usually involve strongly
correlated materials (cuprates, heavy fermions, etc) with localized $d$- or $%
f-$electrons. On the first glance, MgB$_2$ does not seem to belong to any of
such classes. However, it was important to verify experimentally how
reliable are LDA calculations in this compound.

One of the most popular experimental probes of electronic band structure is
angular-resolved photoemission spectroscopy (ARPES), particularly in view of
remarkable progress achieved in the last decade. In spite of the fact that ARPES
probes only a very thin surface layer and is therefore not always
representative of the bulk electronic structure, first experiments\cite%
{ARPES} show an exceptional agreement between the theory and the experiment
in the whole studied energy range. Both $\sigma $ bands and $\pi $ band were
observed along the $\Gamma M$ direction, as predicted by the calculations.
Along $\Gamma K$ direction only one out of the two predicted $\sigma $ bands
was observed; the authors speculated that the single experimental feature in
this region may result from the superposition of the two bands. On the other
hand, the fact that the band in question has different symmetry along the
two measured directions may contribute to the selection rules. In addition,
the analysis of the electronic states centered around the $\Gamma $ point
revealed that this feature originated from a surface electronic state, which
is in good overall agreement between APRES and theoretical results for the
Mg-terminated surface\cite{MGSURF}. Unfortunately, to the best of our
knowledge, surfaces with partial Mg coverage, say, 50\%, were not studied
theoretically, although this is the most likely termination. Possibly even
better agreement can be achieved if such termination will be included in the
calculations.

A classical probe of the Fermi surface properties are quantum oscillations, 
{\it e.g.,} de Haas-van Alphen (dHvA) effect. Such measurements have been
reported \cite{DHVAexp}. Three dHvA frequencies were clearly resolved in
data from Ref.\cite{DHVAexp}, corresponding to two distinct sheets of the
Fermi surface. A comparison of the calculated  frequencies\cite%
{Harima,Rosner,MAZDHVA} with the experimental data shows excellent agreement.
The discrepancies with the theory are less than 300 T which is only 0.2\% of
the area of the hexagonal BZ. The detailed angular dependence of F$_{1}$, F$%
_{2}$ and F$_{3}$ has been calculated in Ref.\cite{Harima} and compares
favorably with the experimental results. The ratio of experimental and
theoretical effective masses provides mass renormalization, presumably of
electron-phonon origin, which appears to be 1.08-1.2 for the inner $\sigma$
cylinder and 0.40 for the $\pi$ sheet. This is to be compared with the
calculated numbers of 1.25\cite{Liu}, 1.57\cite{Kong}, $\sim$1.1 \cite{Choi}%
, and 0.47\cite{Liu}, 0.50\cite{Kong}, $\sim$0.33 \cite{Choi}; a rather good
agreement. Overall, ARPES and dHvA experiments, taken together, fully
support LDA calculations, leaving hardly any room for many body
renormalization of the band masses and velocities, apart from the EPC
renormalization.

It is worth noting that for the $\pi$ orbit it was possible to estimate the
local Stoner enhancement factor. It appears that LDA calculations
underestimate the exchange splitting induced by a magnetic field by about
50\%. The reason for this discrepancy is not clear yet. On the other hand,
electron spin resonance measurements\cite{URBANO,SIMON} found electronic
spin susceptibility of $( 2.0- 2.3 ) \times 10^{-5}$emu/mole, corresponding
to a Stoner renormalization of 50\% {\it less} than calculated\cite%
{NMRantrop,NMReva}.

Since both ARPES and DHVA spectroscopy in MgB$_2$ are described in detail in
other Chapters of this book, we shall refer the reader to those, and will
concentrate in the following on another probe of the electronic structure
near the Fermi level, namely nuclear magnetic resonance (NMR).

NMR spectroscopy measures two  electronic structure related
quantities, the Knight shift, $K$, and the
spin-lattice relaxation rate, $1/T_1T$. The former is related to the
uniform spin susceptibility, the latter to the local susceptibility at a
nucleus. Both are linked to the DOS at the Fermi level, but in an indirect
way involving hyperfine interactions. Therefore extracting reliable
information about the electronic structure is usually possible only if the
corresponding calculations of the hyperfine field are available.

\begin{figure}[tbp] \centerline{
\epsfig{file=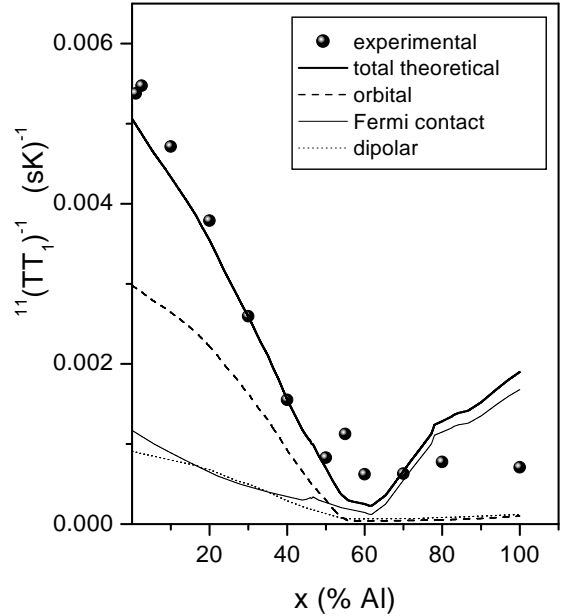,width=0.9\linewidth,angle=0,clip=true}}
\setlength{\columnwidth}{\linewidth} \nopagebreak
\caption{ Boron $^{11}$(1/T$_1$T) for Mg$_{1-x}$Al$_x$B$_2$ as a function of
Al-doping. Lines show the {\it ab initio} calculated plots from Refs.\protect\cite%
{us2,REVMGB2} } \label{NMR} \end{figure}

For MgB$_2$, this is the case. Several experimental groups reported $1/T_1T$ 
\cite{EXP1,EXP2,EXP3} and $K $ for the B site \cite{EXP2,EXP3}, which is of
particular interest because of the role that B states play in
superconductivity. Two groups reported first principles calculations for $%
1/T_1T$ \cite{NMRantrop,NMReva} and for $K$ \cite{NMReva}. Importantly, it
appears that NMR in MgB$_2$ not only probes B electrons, but it also probes
differently $\sigma$ and $\pi$ bands. Indeed, since $\sigma$ bands are
formed by the $p_x$ and $p_y$ states, they can form $p_x\pm p_y$
combinations, which have nonzero orbital moment. One can therefore expect
considerable orbital contribution to the relaxation rate. Indeed,
calculations show \cite{NMRantrop,NMReva} that the orbital mechanism
dominates over the two others, the Fermi-contact and the spin-dipolar,
mechanisms in the spin-lattice relaxation. On the contrary, for the Mg
nucleus the dominant relaxation mechanism is, as usually, the Fermi-contact
interaction, which also dominates the B and Mg Knight shift\cite{NMReva}.

The results of the calculations agree well with the experiment. The
experimental numbers for $1/T_1T$ on $^{11}$B are in a range of $%
(5.6-6.1)\times 10^{-3}/($K$\cdot $sec). Calculations using bare
susceptibility produce numbers of 5.1$\times$10$^{-3}$/(K$\cdot $sec)\cite%
{NMRantrop} and 3.7$\times$10$^{-3}$/(K$\cdot $sec)\cite{NMReva}. This
numbers are subject to many body renormalization. Renormalized values involve
additional assumptions; in Ref.\cite{NMRantrop} the renormalized relaxation
rate was estimated to be 8.1$\times 10^{-3}$/(K$\cdot $sec), while Ref.\cite%
{NMReva} gives a range of(4.3-5.9) $\times$10$^{-3}$/(K$\cdot $sec). As
regards the Knight shift, unfortunately, the spread of the experimentally
obtained values is still too large to allow for a quantitative comparison
with the calculations.

As mentioned above, the NMR relaxation rate is very sensitive to the
relative amount of $\sigma$ and $\pi$ states, which implies a nontrivial
dependence on the filling of the $\sigma$ bands. This was indeed calculated
in Ref.\cite{REVMGB2} for MgB$_2$ doped with Al (whose primary effect is to
fill $\sigma$ hole states). In Ref.\cite{EXP4} the theoretically predicted
in Ref.\cite{REVMGB2} tendencies were experimentally verified for the entire
Mg$_{1-x}$Al$_{x}$B$_{2}$ system of alloys. Very impressive agreement was
obtained (Fig.\ref{NMR}).

\section{Electron-phonon coupling}

\subsection{Standard formulas}

Standard description of the EPC in metals is sometimes referred to as the
Migdal-Eliashberg theory. We are not going to review this theory here, as it
can be found in many excellent texts, but will briefly remind the basic
formulas of this theory. The primary notion of this formalism is that of the
linear EPC vertex, $g_{{\bf k,k+q},\nu}=\langle {\bf k}|dV/dQ_{{\bf q},\nu}|%
{\bf q}\rangle$, where $dV/dQ_{{\bf q},\nu}$ is the derivative of the
crystal potential with respect to the normal phonon coordinate. ${\bf k,k+q}$
stand for the electron wave vectors, and ${\bf q},\nu$ for the wave vector
and the mode index of the phonon whose interaction with the electrons is
being described. In other words, $g_{{\bf k,k+q},\nu}$ is the probability of
an electron to be scattered from the state $|{\bf k}\rangle$ into the state $%
|{\bf k+q}\rangle$ by the phonon (${\bf q},\nu$). Migdal theorem (which
holds for MgB$_2$) states that this vertex is not renormalized by higher
order processes. It does not state, however, as discussed below, that
anharmonic corrections to the phonon spectra or nonlinear vertices like $%
\langle {\bf k}|d^2V/dQ^2_{{\bf q},\nu}|{\bf k+q}\rangle$ are necessaruly
negligible.

$g_{{\bf k,k+q},\nu }$, if properly integrated over all possible virtual
electron-hole pairs, defines the phonon self energy. In particular, its
imaginary part, the phonon linewidth, is given by 
\[
\gamma _{{\bf q},\nu }=2\pi \omega _{{\bf q},\nu }\sum_{{\bf k}}|g_{{\bf %
k,k+q},\nu }|^{2}\delta (\varepsilon _{{\bf k}}-E_{F})\delta (\varepsilon _{%
{\bf k+q}}-\varepsilon _{{\bf k}}-\hbar \omega _{{\bf q},\nu }). 
\]%
In this formula, the right-hand side does not explicitely depend on $\omega
_{{\bf q},\nu }$ (the prefactor cancels  the corresponding factor in $|g_{%
{\bf k,k+q},\nu }|^{2})$. Sometimes a related quantity, the EPC constant for
a given mode, is used: $\lambda _{{\bf q},\nu }=\gamma _{{\bf q},\nu }/\pi
N(E_{F})\omega _{{\bf q},\nu }^{2}$. One may note that this quantity is
strictly zero for optical zone center (${\bf q}=0$) phonons; however, a
related constant can be introduced, $\lambda _{\nu }^{ZZ}=[2N(E_{F})/\omega
_{\nu }]\sum_{{\bf k}}|g_{{\bf k,k},\nu }|^{2},$ and $g_{{\bf k,k},\nu }$ is
obviously related to the deformation potential.

\end{multicols}
When integrated over all phonon modes and corresponding intermediate
electron states, $g_{{\bf k,k+q},\nu }$ defines the electron self-energy,
or mass renormalization $(m^{\ast }/m)_{{\bf k}}$: 
\[
\left( \frac{m^{\ast }}{m}\right) _{{\bf k}}-1=\sum_{{\bf q},\nu }\frac{2}{%
N(E_{F})\omega _{{\bf q},\nu }}{|g_{{\bf k,k+q},\nu }|^{2}\delta
(\varepsilon _{{\bf k}}-E_{F})\delta (\varepsilon _{{\bf k+q}}-\varepsilon _{%
{\bf k}}-\hbar \omega _{{\bf q},\nu })} 
\]

Finally, when integrated over all phonons with given frequency and over
electronic states at the Fermi level, it defines the EPC spectral function,
which determines superconducting properties of a single-gap superconductor, 
\[
\alpha ^{2}F(\omega )=(1/2)\sum_{{\bf q},\nu }{\omega _{{\bf q},\nu }\lambda
_{{\bf q},\nu }\delta (\omega -\omega _{{\bf q},\nu })}, 
\]%
which can be broken into $n\times n$ matrix separating the interband pairing
interaction from the intraband one 
\[
\alpha ^{2}F(\omega )_{ij}=(1/N_i(E_{F}))\sum_{{\bf q},\nu }\sum_{{\bf k}\in i,%
{\bf k+q}\in j}|g_{{\bf k,k+q},\nu }|^{2}\delta (\varepsilon _{{\bf k+q}%
}-\varepsilon _{{\bf k}}-\hbar \omega _{{\bf q},\nu })\delta (\omega -\omega
_{{\bf q},\nu }), 
\]%
where $i,j$ label different electronic bands or group of bands, $e.g.,$ $%
i=\sigma ,\pi $.
\begin{multicols}{2}

\subsection{First principle calculations}

In the first publication~\cite{us} following the discovery of SC in MgB$%
_2 $ the strength of the EPC was estimated and it was suggested that MgB$_2$
is a standard BCS superconductor, where coupling with the B phonons is the
driving force for superconductivity. A substantial B, but small Mg isotope
effects were predicted. Both predictions were confirmed by the experiment%
\cite{Budko,Mgiso}. The relevant phonons were soon identified in Ref.\cite{An} as
two optical $E_{2g}$ modes, which was confirmed by subsequent full-scale
calculations of EPC.

Because of pronounced dissimilarity between different electron groups and
different phonon modes it is unavoidable for understanding superconductivity
in MgB$_{2}$ to calculate EPC spectral function $\alpha ^{2}F(\omega )$
including all bands and all phonons on the same footing. By now, at least
four groups have claimed to have done this from the first principles\cite%
{Liu,Kong,Choi,Bohnen}. Three of these \cite{Liu,Choi,Bohnen} were based on
pseudopotential band structure calculations; one \cite{Kong} utilized a
full-potential LMTO method. Three used the linear response formalism to compute
phonon spectra and electron-phonon matrix elements\cite{Liu,Kong,Bohnen}; one%
\cite{Choi} was based on frozen phonon calculations at several high-symmetry
point with a subsequent interpolation onto a finer mesh. The last work also
used an anharmonic correction to the phonon frequencies, which the other
three works did not include (Ref.\cite{Liu} provided a rough estimate of the
effect). The results are compared in Table \ref{phon}, and the $E_{2g}$
frequencies are compared with selected frozen phonon calculations.

The last two columns in Table \ref{phon} show {\it isotropic}, or
thermodynamic EPC constant; as discussed later, it is probably not directly
relevant to superconductivity, but it defines the average electronic mass
renormalization, and thus the renormalization of specific heat. The latest
exteriments\cite{lastheat,HEATEXP,MACHIDA} (the latter two on single
crystals), reported for the electronic specific heat coeffiient the values
of $\gamma=$2.5, 2.3, and 3.5 mJ/mole$\cdot$K$^{2}$, respectively (the
discrepancy may be partially related to different temperature ranges used in
fitting). The unrenormalized DOS (Table \ref{phon}) corresponds to $\gamma=$%
1.67 mJ/mole$\cdot$K$^{2}$, yielding $\lambda$ from 0.4 to 1.1. While
clearly inconclusive, these numbers are equally consistent with all entries
in the Table \ref{phon}.
\end{multicols}
\begin{table}
\caption{Electron-phonon calculations and selected calculations of other
relevant parameters, as reported in the literature.}
\label{phon}%
\begin{tabular}{c|cccccc}
& $\omega _{E_{2g}}^{harm}$, cm$^{-1}$ & $\omega _{E_{2g}}^{anharm}$, cm$%
^{-1}$ & $\omega _{\log }$, cm$^{-1}$ & $N(E_{F})$, st./Ry spin & $\lambda
^{harm}$ & $\lambda ^{anharm}$ \\ 
\tableline Ref\cite{Kong} & 540$^{(a)}$ &  & 504 & 4.83 & 0.87 &  \\ 
Ref\cite{Bohnen} & 536 &  & 487 &  & 0.73 &  \\ 
Ref\cite{Liu} &  &  & 450 & 4.83 & 0.77 & 0.70 \\ 
Ref\cite{Choi} & 506 & 612 & 479 & 4.83 & 0.73 & 0.61 \\ 
FPLAPW\cite{Liu} & 536$^{(a)}$ & 590 &  & 4.80 &  &  \\ 
&  &  &  &  &  & 
\end{tabular}%
$^{(a)}$updated results with a better $k$-point convergence (J. Kortus,
private communication)
\end{table}

\begin{multicols}{2}
As regards the EPC there is a noticeable discrepancy between different
calculations, despite an overall agreement. Part of that may be due to
different band structure techniques, but the difference is too large to be
ascribed to the band structure difference alone (note nearly perfect
agreement between the calculated DOS in Table \ref{phon}). At least part of
the difference comes from the difference in the calculated phonon
frequencies. Direct calculations of the phonon frequencies by the
frozen-phonon technique are generally more reliable and less sensitive to
the the size of the basis set than linear response methods. All-electron
calculations are usually more reliable than pseudopotential calculations.
Therefore we included in the Table the results of full-potential LAPW
calculations. In view of high sensitivity to the phonon spectra, the fact
that only a handful of high-symmetry points were treated from the first
principles in Ref.\cite{Choi} is a weak point of this work.

However, the differences in the phonon spectra do not explain the
discrepancy in the value of the calculated EPC constants. To understand
where this discrepancy possibly originates, let us note that if the $\sigma $%
-band Fermi surfaces were ideal cylinders (which they nearly are), the EPC
for the $E_{2g}$ phonons would have two Kohn-like divergencies\cite{oleg}.
Indeed, it is easy to show that in this case the partial EPC constant for a $%
E_{2g}$ phonon with a wave vector $q$, $\lambda _{q}$, is given by the
expression 
\[
\lambda _{q}\approx \frac{\langle g^{2}\rangle }{2\pi E_{F}\omega _{q}x\sqrt{%
1-x^{2}}} 
\]%
where $\langle g^{2}\rangle $ is the average EPC matrix element, $\omega
_{q}\approx \omega _{0}$ is the phonon frequency, and $x=q/2k_{F}$. Note
that $\lambda _{q}$ is inversely proportional to the Fermi energy, and
therefore to the number of phonons with $q<2k_{F}$, so that the total $%
\lambda $ given by the sum over all $E_{2g}$ phonons does not depend on the
size of the Fermi surface. This is, of course, simply a reflection of the
fact that the DOS of a 2D band does not depend on the Fermi energy, and the
total $\lambda $ is, essentially, just total DOS times the average squared
EPC matrix element. 
\begin{figure}
\centerline{
\epsfig{file=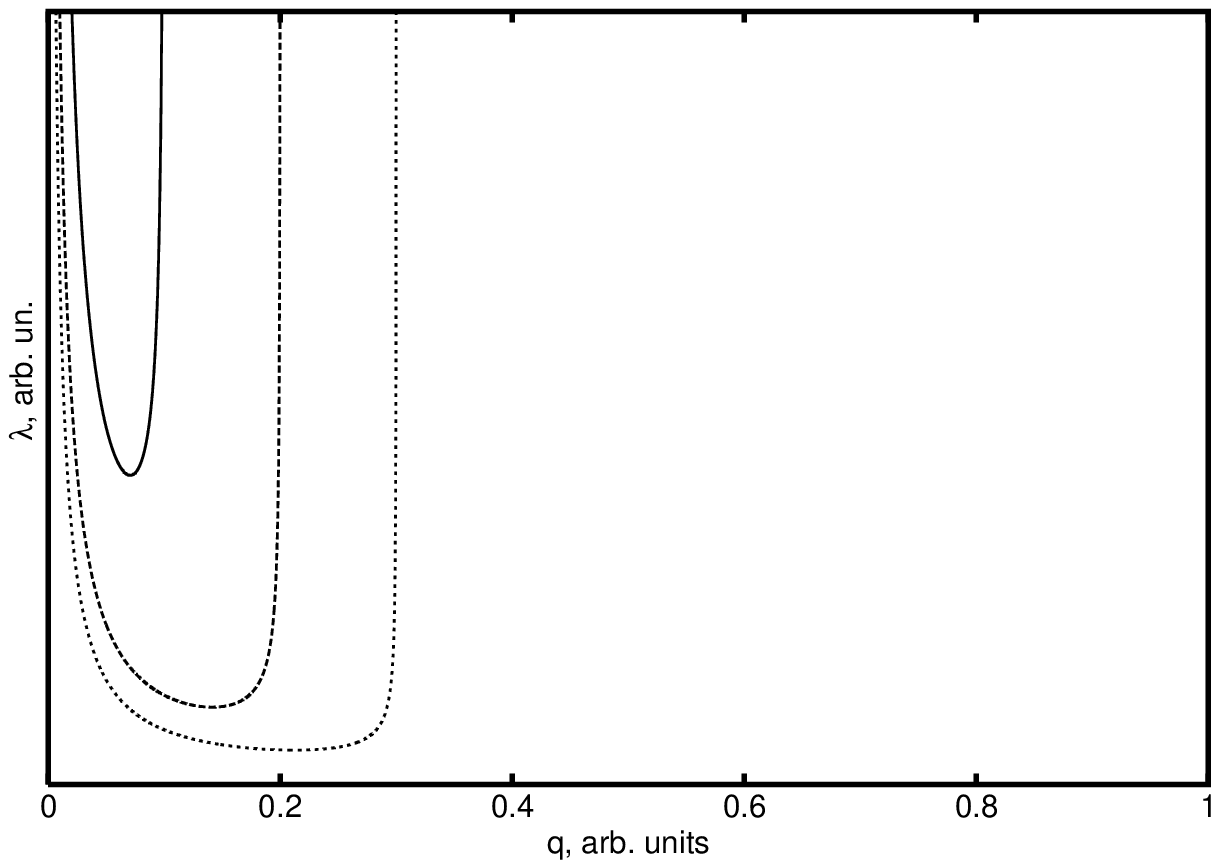,width=0.9\linewidth,clip=true}}
\setlength{\columnwidth}{\linewidth} \nopagebreak
\caption{Dependence of the partial EPC constant on the phonon wave vector
for a cylindrical Fermi surface. Note singularities at small $q$ and at $%
q=2k_{F}$. Three curves correspond to three different $k_{F}$, but all
integrate to the same total $\protect\lambda $.}
\label{sing}
\end{figure}

This function is plotted in Fig.\ref{sing}. Essentially, in the calculations
like Refs.\cite{Liu,Kong,Choi,Bohnen} one needs either to integrate these
singularities numerically or apply to them a special analytical treatment.
The first approach was employed in Refs.\cite{Kong,Bohnen,Choi}. In
particular, in Ref.\cite{Kong} a special care was taken to assure that the
singularity was properly integrated. In Ref. \cite{Liu} the small $q$
singularity was treated analytically; but not the high-$q$ one. Later
estimates [I.I. Mazin, unpublished] show that the discrepancy between Refs.\cite%
{Kong} and \cite{Liu} is substantially reduced when the high-$q$ singularity
is treated analytically as well, although the total $\lambda$ remains
slightly smaller that in Ref.\cite{Kong}.

\subsection{Phonon renormalization, anharmonicity, and nonlinear coupling}

In this Section we will address several seemingly unrelated, but in fact
strongly connected issues. As mentioned above, calculated frequencies of the 
$E_{2g}$ phonon show strong anharmonicity\cite{Choi}. At the same time,
calculations show this phonon to soften abruptly around $q<2k_F$, where $k_F$
is the Fermi vector for the $\sigma$ bands\cite{Kong,Bohnen}. Finally, it
was noticed that the matrix elements for quadratic EPC, $g^{quad}=\langle|%
\delta^2 V/\delta Q^2|\rangle$ are anomalously large compared with that for
the linear coupling, $g^{lin}= \langle|\delta V/\delta Q |\rangle $\cite%
{Liu,Yildirim}. 
\begin{figure}[tbp]
\centerline{
\epsfig{file=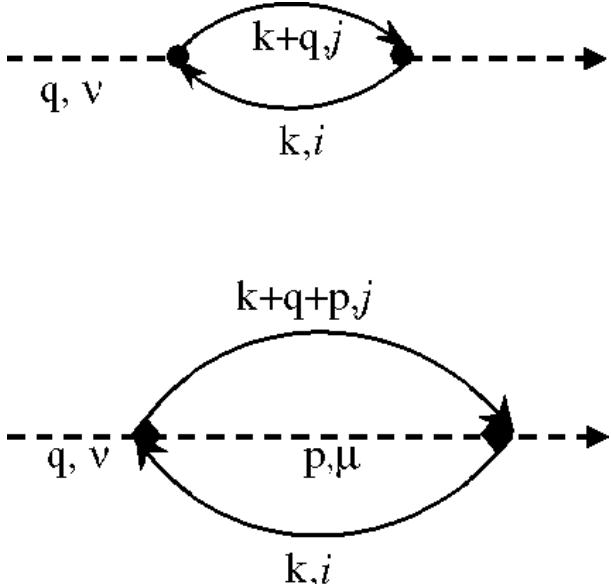,width=0.9\linewidth,angle=-90,clip=true}}
\setlength{\columnwidth}{\linewidth} \nopagebreak
\caption{Examples of the processes contributing to the phonon self energy in
the linear (top) or quadratic (bottom) approximations for the EPC.}
\label{diag}
\end{figure}

To understand these effects we should recall that in the linear coupling
regime the effect of the electronic screening on the phonon self-energy (Fig.%
\ref{diag}, top) is defined by the same process that determines the
contribution of the corresponding phonon to the total superconduction EPC
constant. Indeed, the imaginary part of the phonon self energy (phonon
linewidth) is related to $\lambda _{q}$ as $\gamma _{q}=\pi N(E_{F})\omega
_{q}^{2}\lambda _{q}$. At the same time, the real part of the same
self-energy defines phonon softening. Only the phonons with $q<2k_{F}$ can
couple with the $\sigma -$electrons, therefore they and only they become
screened and softened by them. For a zone-center phonon, there is a
quanitative measure of this softening\cite{Rodriguez}: 
\begin{equation}
\Delta \omega ^{2}=-4\omega \langle g^{2}\rangle N(E_{F}),  \label{soft}
\end{equation}%
where the right-hand side does not depend on $\omega $. This quantity was
calculated in Ref. \cite{Liu} to be\cite{0.51} approximately $ 2\times 0.51\omega
^{2}\approx 1.02\ast 503^{2}$ cm$^{-2}$. This corresponds to a bare
frequency of 715 cm$^{-1}$. In the same work, the frequency of the $E_{2g}$
phonon away from the $\Gamma $ point was calculated to be around 640 cm$%
^{-1} $. Softening from 715 to 640 cm$^{-1}$ must therefore be coming from
the screening due to the $\pi $-electrons. Given high sensitivity of phonon
frequencies to the $k$-mesh convergency, one can probably say that first
principle calculations give a softening due to $\sigma $-electrons of 75-100
cm$^{-1}$.

Eq. \ref{soft} is based on the linear approximation, that is, EPC is
proportional to the first derivative with respect to the phonon coordinate.
This is, however, not an easily justifieable approximation in case of MgB$_2$%
: as we saw above, the second-order EPC vertex, $g^{quad}$, is anomalously
large. In this case one has to consider in the phonon renormalization
processes corresponding to creation/annihilation of an electron-hole pair,
associated with emission/absorption of two $E_{2g}$ phonons, as illustrated
in Fig.\ref{diag} (bottom). Note that the corresponding diagrams are
temperature dependent, therefore producing intrinsically anharmonic phonons,
as observed in the frozen phonon calculations. Quadratic EPC is a long known
phenomenon (see, e.g., Refs.\cite{Hui,Heid,Crespi}), although most authors
concentrated on its effect on superconductivity and mass renormalization,
rather than on phonon frequencies.

In order to gain a better insight into the interrelation between the
anharmonicity, quadratic coupling, and frozen phonons, let us look for the
reason for the anomalously large quadratic vertex. One can conviniently
write the dispersion of the two $\sigma$-bands as $\epsilon_{{\bf k}}= u_{%
{\bf k}}\pm v _{{\bf k}} $, where both $u$ and $v$ are quadratic functions
of $k$, and $v=0$ at $k=0$. The function $u$ describes the average
dispersion neglecting hybridization between the two bands, while $v$
describes the hybridizations. Both functions depend on the frozen phonon
coordinate, but in a different way: for a given point ${\bf k}$, $u$ is an
odd function of the phonon coordinate, $du _{{\bf k}}/dQ \neq 0$; however,
any symmetry lowering increases hybridization between the two $\sigma$
bands, therefore $v$ is an even function of $Q$, $dv_{{\bf k}}/dQ=0$, $d^2v_{%
{\bf k}}/dQ^2\neq 0$. At the $\Gamma$ point $du /dQ =0$, therefore only
nonlinear coupling remains; when going away from the $\Gamma$ point, a
nonzero linear component appears (which is responsible for a large
calculated EPC in Refs.\cite{Kong,Bohnen,Liu,Choi}), and quadratic coupling
gradually vanishes. Correspondingly, the smaller is the number of holes in
the $\sigma$ band the stronger are anharmonic effects in the phonon
frequency.

The same can be seen from the point of view of the frozen phonon
calculations. These amount to calculating total energy of a crystal with
fixed ionic displacement comparable with, or smaller than the amplitude of
the zero-point oscillations. This energy remains more or less harmonic as
long as the frozen displacement does not incur any change in the Fermi
surface topology. This ``critical'' displacement becomes smaller when the $%
\sigma$-pockets get filled, therefore yielding more and more anharmonic
phonons, in perfect agreement with the reasoning above.

The interrelated nonlinearity and anharmonicity have competing effects on
superconductivity. Anharmonic hardening of the phonon reduces effective EPC
constant (Table\ref{phon}), while two-phonon exchange provides an additional
contribution to $\alpha^2F(\omega)$ at frequencies roughly twice the
frequency of the $E_{2g}$ phonon. The latter effect was never reliably
calculated. Estimates of Yildirim $et$ $al$\cite{Yildirim} allow one to
assume that nonlinear EPC increases the coupling constant for $\sigma$ bands
by at least 5\%, although this is probably the lower estimate.

\section{Multiband effects in superconductivity}

Already in the first months after the discovery of superconductivity in MgB$%
_2$ experiments appeared that were not consistent with a conventional strong
coupling superconductivity scenario. It was observed that the critical field 
\cite{shulga}, specific heat\cite{junod} and tunneling\cite{Bascones}
measurements are easier to explain if two gaps are assumed instead of one.
Liu $et$ $al$\cite{Liu} proposed, based on electronic structure and EPC
calculations, that there are, in fact, two distinctive gaps associated with $%
\sigma$- and $\pi$-Fermi surfaces. This ``two-gap'' model gained popularity,
and it became clear that the EPC calculations needed to be performed
separately for the two sets of bands.

With this in mind, the results of Ref.\cite{Liu} and subsequently of Ref\cite%
{Kong} were broken in a 4x4 EPC coupling matrix, as well as in a 2x2 matrix
(Table\ref{lam}). Ref.\cite{Choi} does not report the corresponding 2x2
matrix, but it can be reasonably accurately restored from the figures in
that paper. Detailed calculations\cite{Choi} show that in the ideally clean
limit the variation of the order parameter, apart from the $\sigma -\pi $
difference, are less than 10\%. As discussed below, such a variation cannot
exist in real sample even with an extremely small impurity concentration,
therefore it is of little interest to use more than 2x2 EPC matrix in any
physically relevant discussion.

\subsection{General Theory}

The famous BCS formula is derived in the assumption that the pairing
amplitude (superconducting gap, order parameter) is the same at all points
on the Fermi surface. The variational character of the BCS theory makes one
think that giving the system an additional variational freedom of varying
the order parameter over the Fermi surface should always lead to a higher
transition temperature. This problem was solved first in 1959 by Matthis,
Suhl, and Walker\cite{Suhl} and by Moskalenko\cite{Moskal}. The general
solution was given later by several authors (probably in the most developed
form by Allen and collaborators\cite{Allen78}), and for our purpose can be
written as 
\begin{equation}
\Delta ({\bf k})=\int \Lambda ({\bf k,k}^{\prime })\Delta ({\bf k}^{\prime
})F[\Delta ({\bf k}^{\prime }),T]d{\bf k}^{\prime },
\end{equation}%
where summation over {\bf k} implies also summation over all bands crossing
the Fermi level. The matrix $\Lambda $ characterizes the electron-phonon
interaction, and the temperature dependence is given by the function $%
F=\int_{0}^{\omega _{D}}dE\tanh (\frac{\sqrt{E^{2}+\Delta ^{2}}}{2T})/\sqrt{%
E^{2}+\Delta ^{2}}$ . For the purpose of this paper it suffices to use the
discrete (also called disjoint) representation, where it is assumed that the
order parameter $\Delta $ varies little within each sheet of the Fermi
surface, while differing between the different sheets: 
\begin{equation}
\Delta _{i}=\sum_{j}\Lambda _{ij}\Delta _{j}F(\Delta _{j},T),
\end{equation}%
where $i,j$ are the band indices and $\Lambda $ is an {\it asymmetric}
matrix related to the {\it symmetric }matrix of the pairing interaction, $%
\Lambda _{ij}=V_{ij}N_{j}$, where $N_{i}$ is the contribution of the $i$-th
band to the total DOS. It can be shown that in the BCS weak coupling limit
the critical temperature is given by the standard BCS relation, $%
kT_{c}=\hbar \omega _{D}\exp (-1/\lambda _{eff}),$ where $\lambda _{eff}$ is
the largest eigenvalue of the matrix $\Lambda .$ The ratios of the
individual order parameters are given by the corresponding eigenvector. Note
that although the matrix $\Lambda $ is not symmetric, its eigenvalues are
the same as those of the symmetric matrix $\sqrt{N}V\sqrt{N}$.

\begin{table}[tbp] 
\caption{2x2 EPC matrices in different calculations.}
\label{lam} \begin{tabular}{|c|c|c|}
Ref.\cite{Liu} & Ref.\cite{Golheat} & Ref. \cite{Choi}$^{a}$ \\ 
\tableline $
\begin{array}{cc}
0.96 & 0.17 \\ 
0.23 & 0.29
\end{array}
$ & $%
\begin{array}{cc}
1.02 & 0.16 \\ 
0.21 & 0.45%
\end{array}%
$ & $%
\begin{array}{cc}
0.78 & 0.11 \\ 
0.15 & 0.21%
\end{array}%
$ \\ 
&  & 
\end{tabular}
$^{a}$ Obtained by integrating $\lambda ({\bf k,k^{\prime }})$ distribution
plots from Ref.\cite{Choi}.
\end{table}

The mass renormalization parameters for each band can be constructed from
the matrix $\Lambda $: $\lambda _{i}=\sum_{j}\Lambda _{ij}. $ These $\lambda
_{i}$ define, among other things, the de Haas-van Alphen thermal masses.
Finally, the renormalization of the specific heat is given by the weighted
average of $\lambda _{i},$ $\bar{\lambda}=\sum_{i}N_{i}\Lambda
_{ij}/N_{tot}=\sum_{ij}N_{i}\Lambda _{ij}N_{j}/N,$ which is also the
``Eliashberg'' coupling constant determining the superconductivity in the
isotropic limit, where all order parameters are constrained to be the same.
One can show that $\bar{\lambda}\leq \lambda _{eff},$ the equality being
achieved when and only when all elements of the $V$ matrix are the same (the
relative magnitude of $N_{i}$ is irrelevant). Physically this result is
obvious: the BCS theory can be formulated as a variational theory. Therefore a
bigger energy gain, and a higher critical temperature, can be achieved if
more variational freedom is provided, $e.g.,$ by allowing different order
parameters in the different bands.

\subsection{Impurity scattering}

In this Section we will ouline nontrivial effects related to impurity
scattering in a multigap superconductor. The discussion will mostly follow
Ref.\cite{golubov}, where more details can be found. In the Born
approximation, and close to $T_{c},$ the problem can be solved analytically.
It appears that nonmagnetic impurities suppress superconductivity in much
the same way, as magnetic ones do in a regular superconductor, however, only
the {\it inter}band impurity scattering has a pair-breaking effect. In the
weak nonmagnetic scattering limit, for two bands, the $T_{c}$ suppression is%
\begin{equation}
\frac{\delta T_{c}}{T_{c}}=-\frac{\pi \gamma _{12}}{8kT_{c}}\frac{(\Delta
_{1}-\Delta _{2})(\Delta _{1}N_{2}-\Delta _{2}N_{1})}{(\Delta
_{1}^{2}+\Delta _{2}^{2})N_{2}},
\end{equation}%
where $\gamma _{12}\equiv \gamma _{21}N_{2}/N_{1}$ is the interband
scatering rate. Note that the $T_{c}$ suppression is linear in $\gamma
_{12}. $ This formula also gives us a clue about what is a weak and what is
a sttrong scattering in the specific case of MgB$_{2}:$ small scattering is
when $\gamma _{12}\ll (\delta \Delta ^{2}/\bar{\Delta}^{2})T_{c},$ where $%
\delta \Delta $ is the variation of the gap between the bands, and $\bar{%
\Delta}$ is the average gap. The ratio of the $\sigma -$ and the $\pi -$
band gaps is, experimentally and theoretically, of the order of 3. The
densities of states are comparable. Therefore a $T_{c}$ suppression of 1 K
would require an interband scattering rate of the order of 1 meV. It is a
fortunate and rather unexpected coincidence that the symmetry of the
electronic states conspire in such a way as to make the interband scattering
rate quite small even in rather dirty samples\cite{imp}. Only because of
this conspiracy we are actually able to observe two distinctive gaps in this
compound.

On the other hand, the variation of the gap within each of the two band
systems, calculated in Ref.\cite{Choi}, which is of the order of 7\%, cannot
survive a $\sigma -\sigma $ impurity or phonon scattering stronger than $%
\sim 0.01$ meV, and therefore is unobservable in samples of any imaginable
quality.

In the strong interband scattering limit a complete isotropization of all
Fermi surfaces takes place. This limit is achieved\cite{golubov} when the
interband scattering rate becomes larger than the relevant phonon frequency,
in our case, 600 cm$^{-1}\approx 75$ meV. Then the two gaps merge to one, the
isotropic BCS gap, and the critical temperature drops to it isotropic value.
Strong coupling calculations of Ref.\cite{Choi} predict the latter to be
around 19 K. Indeed, recent experiments on irradiated samples\cite{Junod-irr}
demonstrated a reduction of the gap ratio by 40\%, accompanied by a $T_{c}$
reduction by 22\%. One should note, however, that the results of Ref.\cite%
{Junod-irr}, while qualitatively consistent with the prediction of the
two-band model, quantitatively do not agree with them. Similar results were reported in Ref.
\cite{canf}.

\subsection{Strong coupling and Coulomb pseudopotential}

It is relatively straightforward to extend the theory of multiband
superconductivity beyond the weak coupling BCS model\cite{Allen78}.
Qualitatively one can easily understand the main effect of the strong
coupling by recalling the McMillan equation:%
\begin{equation}
kT_{c}=\frac{\hbar\omega _{\log }}{1.2}\exp \left[ \frac{-1.02(1+\lambda )}{%
\lambda -\mu ^{\ast }(1+0.62\lambda )}\right] .
\end{equation}%
Qualitatively, this equation can be understood as renormalized BCS equation, 
$kT_{c}=\hbar\omega _{ph}\exp [ -1/(\lambda -\mu ^{\ast })] ,$ where $\omega
_{ph}=\omega _{\log }/1.2,$ and the mass renormalization has been applied to 
$\lambda $, $\lambda \rightarrow \lambda /(1+\lambda ).$ We already know
that the multiband version of the BCS equation differs from this in that $%
\lambda $ is substituted by an effective $\lambda _{eff},$ the largest
eigenvalue of the matrix $\Lambda .$ The effect of Coulomb repulsion,
introduced in the BCS model $via$ the Coulomb pseudopotential $\mu ^{\ast },$
is likewise introduced in its multiband version $via$ the {\it matrix} $\mu
_{ij}^{\ast }$. The multiband analog of the McMillan equation is, therefore, 
\begin{equation}
kT_{c}=\frac{\hbar\omega _{\log }}{1.2}\exp \left[ \frac{-1}{(\lambda -\mu
^{\ast })_{eff}}\right] ,
\end{equation}%
where $(\lambda -\mu ^{\ast })_{eff}$ is defined as the maximum eigenvalue
of the matrix%
\begin{equation}
\Lambda _{ij}^{eff}=\frac{\Lambda _{ij}-\mu _{ij}^{\ast
}(1+0.62\sum_{n}\Lambda _{in})}{1+\sum_{n}\Lambda _{in}}.
\end{equation}%
This expression gives the results very close to the full solution of the
multiband Eliashberg equations.

The Coulomb pseudopotential matrix is not a constant, as it is sometimes
believed\cite{Choi}. First of all, already the {\it bare} pseudopotential
matrix, $\mu _{ij},$ is not uniform. Indeed, it is formally defined as $%
\left\langle \left\langle V_{C}\right\rangle \right\rangle _{ij}N_{j}$
(where $V_{C}$ is the screened Coulomb interaction, and the averaging is
over the corresponding Fermi surfaces), and as had been noticed, for
instance, by Agtergerg {\it et al} in another compound\cite{ARS}, when
different bands have different orbital character, the Coulomb matrix
elements between these bands are suppressed compared to intraband matrix
elements. Jepsen and Andersen\cite{comment2} estimated this effect, using
the tight-binding LMTO method and found the ratio between $\left\langle
\left\langle V_{C}\right\rangle \right\rangle _{\sigma \sigma }$, $%
\left\langle \left\langle V_{C}\right\rangle \right\rangle _{\pi \pi }$ and $%
\left\langle \left\langle V_{C}\right\rangle \right\rangle _{\sigma \pi }$
to be $\approx $3:2.5:1. Furhermore, any anisotropy in {\it bare}
pseudopotential is further enhanced in the renormalized $\mu _{ij}^{\ast }.$
In the one-band case $\mu $ is renormalized as $\mu ^{\ast }=[\mu /(1+\mu
\log (W/\omega _{\log }))],$ where $W$ is a {characteristic electronic
frequency} (of the order of the bandwidth or plasma frequency). For a
multiband case we have a matrix equation, which is a natural extension of
the standard procedure\cite{schrieffer} 
\begin{equation}
\mu _{ij}^{\ast }=\mu _{ij}-\sum_{n}\mu _{in}\log (W_{n}/\omega _{c})\mu
_{nj}^{\ast }.  \label{eq:mustar}
\end{equation}%
It is easy to show that renormalization enhances any nonuniformity in $\mu ;$
indeed, assuming $\mu _{\sigma \sigma }=\mu _{\pi \pi }=\alpha \mu _{\sigma
\pi }$, ($\alpha >1)$, and $\mu _{\sigma \sigma }\log (W_{\sigma }/\omega
_{c})=\mu _{\pi \pi }\log (W_{\pi }/\omega _{c})=L,$ we obtain $\alpha
^{\ast }=\alpha +(\alpha -1/\alpha )L.$ From the ratios of $\left\langle
\left\langle V_{C}\right\rangle \right\rangle $'s above, $\alpha \sim 2.3,$
and $L$ for MgB$_{2}$ is of the order of 0.5 - 1, so for $\mu _{ij}^{\ast }$
it holds that $\mu _{\sigma \sigma }^{\ast }=\mu _{\pi \pi }^{\ast }\sim
4\mu _{\sigma \pi }^{\ast },$ 

The fact that the matrix $\mu _{ij}^{\ast }$ is approximately diagonal is of
utmost importance. Various calculations \cite{Liu,Kong,Choi} differ in
details, but all agree that the interband electron-phonon coupling constant
is 0.15-0.2. Since the order parameter in the $\pi $ band is induced by the $%
\sigma $ band (except for the very low temperature), if a Coulomb repulsion
offsets most of the interband coupling, the induced gap becomes vanishingly
small. If $\mu _{\sigma \pi }^{\ast }$ were of the order of $\mu _{ii}^{\ast
}\approx 0.1,$ the gap ratio $\Delta _{\sigma }/\Delta _{\pi }$ would be
much larger than the observed ratio of approximately 3. It is worth
mentioning that this is in direct contradiction with a popular misconception
that \textquotedblleft the superconducting properties of MgB$_{2}$ are not
very sensitive to $\mu ^{\ast }$\textquotedblright \cite{Choi}; they are not
only in the one-band picture. To demonstrate this, we performed\cite%
{comment2} 2x2 Eliashberg calculations using the electron-phonon interaction
from Ref. \cite{Choi}. Although the authors of Ref.\cite{Choi} do not break
down their results for the electron-phonon coupling in a 2-band form, which
would have made them easier to analyze, one can find the 2x2 matrix
corresponding to their calculations by integrating the $\lambda ({\bf k,k}%
^{\prime })$ distribution  depicted in their graphs (Table \ref{lam}). It
appeared that with $\mu ^{\ast }(\omega _{c})$=0.12, used in Refs.%
\cite{Choi}, the ratio $\Delta _{\sigma }/\Delta _{\pi }$ at
zero temperature is 4.3 and the critical temperature $T_{c}=45$ K. On the
other hand, calculations with a {\it diagonal} matrix, $\mu _{\sigma \pi
}^{\ast }=0,$ $\mu _{\sigma \sigma }:\mu _{\pi \pi }=N_{\sigma }:N_{\pi },$
produced $T_{c}=39$ K and $\Delta _{\sigma }/\Delta _{\pi }=3.1${\it .} 

\subsection{Normal transport}

A closer look at normal transport in MgB$_2$ reveals several phenomena which
are hard to understand. First, there is a severe violation of the
Matthiessen rule: samples with large residual resistivity tend to have much
stronger temperature dependence of the resistivity than ``clean'' samples.
Second, optical conductivity does not seem to obey the Drude-Lorenz law; if
one attempts a Drude-Lorenz fit to experimental spectra, the extracted
plasma frequency is 5 times smaller than expected. Many researchers believe
that these problems are due to extrinsic effects like grain boundaries.
While future experiments will clarify this matter, it interesting to
observe that multiband effects can actually explain such observations rather
easily.

The theory of multiband effects in electric transport has been developed by
Allen and co-workers\cite{PBA}. One important qualitative statement can be
made upfront: since the kinetic equation in a metal can be solved
variationally with respect to the electric conductivity, giving a
variational freedom for different bands to change their distribution
functions separately should always result in an increase of the
conductivity. In other words, while in the one-band theory the
superconducting, the thermodynamic, and the transport PPC constants are
usually similar (the first two being identical), in the multiband theory the
former is always larger than in the corresponding one-band scenario, and the
latter is always smaller. Quantitatively, one can write down the following
formulas:%
\begin{eqnarray}
\sigma &=&e^2\sum_{ij}(\rho ^{-1})_{ij} \\
\rho _{ij} &=&t_{ij}/ [\sum_{{\bf k } }v_{i{\bf k}}^{2}\delta (\varepsilon
_{i{\bf k}})] [\sum_{{\bf k } }v_{j{\bf k}}^{2}\delta (\varepsilon _{j{\bf k}%
})] \\
t_{ij} &=&t_{ij}^{out}-t_{ij}^{in} \\
&=&\delta _{ij}\sum_{{\bf kk}^{\prime }n}P_{i{\bf k},n{\bf k}^{\prime }}v_{i%
{\bf k}}^{2}\delta (\varepsilon _{i{\bf k}})\delta (\varepsilon _{n{\bf k}%
^{\prime }}) \\
&&-\sum_{{\bf kk}^{\prime }}P_{i{\bf k},j{\bf k}^{\prime }}v_{i{\bf k}}v_{j%
{\bf k}^{\prime }}\delta (\varepsilon _{i{\bf k}})\delta (\varepsilon _{j%
{\bf k}^{\prime }}),
\end{eqnarray}%
where $v_{i{\bf k}}$ is the electron velocity along the direction of the
current. The physical meaning of these formulas is just that of the parallel
conductors formula, each element of the matrix $\rho ^{-1}$ representing a
separate conductor. If the scattering probability $P_{i{\bf k},j{\bf k}%
^{\prime }}$ is reasonably isotropic, averaging over the Fermi surface
renders $t_{ij}^{in}$ very small. Neglecting it, and using the standard
expressions for the phonon-limited and impurity parts of $P_{i{\bf k},j{\bf k%
}^{\prime }},$ we have, for two bands,%
\begin{eqnarray}
1/\rho _{{\rm DC}}(T) &=&\frac{1}{4\pi }\left( \frac{\omega _{{\rm pl\,}%
\,\pi }^{2}}{\Gamma_{\pi }(T)}+\frac{\omega _{{\rm pl\,}\,\sigma }^{2}}{%
\Gamma_{\sigma }(T)}\right) ,  \label{rho} \\
\Gamma_{\sigma }(T) &=&\gamma _{\sigma \sigma }+\gamma _{\sigma \pi }+\frac{%
\pi }{T}\int_{0}^{\infty }d\omega \frac{\omega }{\sinh ^{2}(\omega /2T)} 
\nonumber \\
&\times &\left[ \alpha _{{\rm tr}}^{2}(\omega )F_{\sigma \sigma }(\omega
)+\alpha _{{\rm tr}}^{2}(\omega )F_{\sigma \pi }(\omega )\right] ,  \nonumber
\end{eqnarray}%
where the plasma frequencies are defined in their usual way for each
Cartesian direction. The disparity of the two band systems appears here in a
trivial way, through different $\omega _{{\rm pl}}^{2},$ and in a
non-trivial way, through different $\alpha _{{\rm tr}}^{2}F(\omega )$ and
different $\gamma .$ As described elsewhere in this Chapter, the
electron-phonon scattering is much stronger in the $\sigma \sigma $ channel
that in the other channels, and the impurity scattering is essentially
always small in the interband channel; furthermore, it is often much
stronger in the $\pi \pi $ channel than in the $\sigma \sigma $ channel. The
smallness of the interband impurity scattering is essential for the two-gap
superconductivity; the sample-dependence of the {\it intraband }$\gamma ,$
especially of the $\gamma _{\pi \pi },$ is important for the understanding
of the temperature dependence of the normal resistivity. Indeed, it is
usually assumed that the impurity scattering is, in the first approximation,
irrelevant for the temperature dependence of the resistivity. It is not
necessarily true in a two-band system.

To start with, let us consider a very clean sample, $\gamma _{ij}=0.$ The
in-plane conductivity at $T=0$ is defined by both bands, but mostly by the $%
\pi $ band, because it has a larger plasma frequency. The out-of-plane
conductivity, of course, is defined by the $\pi $ band only. Closer to room
temperature the contribution of the $\sigma $ band becomes smaller and
smaller, because of the strong EPC scattering in this band. Eventually, the
high-$T$ behaviour is dominated by the $\pi $ band with its small EPC
constant. Temperature dependence at the high temperature (above room
temperature) is therefore weak. Let us now consider a dirty sample with $%
\gamma _{\pi \pi }\gg \gamma _{\sigma \sigma }\gg \gamma _{\sigma \pi }.$
Because of the strong impurity scattering, $\pi -$ electrons contribute very
little to superconductivity, so the temperature dependence is defined
entirely by the EPC in the $\sigma $ bands - and thus is strong. For a more
detail discussion of these issues we refer the reader to the paper\cite{imp}.

Similar effects are expected in optical conductivity; the relevant formulas
differ from Eq.\ref{rho} only in the sense that a frequency dependence of
the EPC scattering should be taken into account in the usual way, and in the
first line $\Gamma_{i}(T)$ should be substituted by $\Gamma_{i}(T)-i\omega .$
Nontrivial effects may be expected in the \textquotedblleft
dirty\textquotedblright\ regime\cite{Kuzmenko,comment} $\gamma _{\pi \pi
}\gg \gamma _{\sigma \sigma }\gg \gamma _{\sigma \pi }.$ In this regime the
Drude peak in optical conductivity that stems from the $\pi $ - electrons
broadens, possibly beyond recognition, and manifests itself merely as a flat
background. Analyzing such a conductivity will uncover only one Drude peak,
the one due to $\sigma $ - electrons, with a much reduced spectral weight,
compared to the total plasma frequency. Moreover, if $\gamma _{\sigma \sigma
}\lesssim \omega _{ph}\approx 70$ meV, where $\omega _{ph}$ is the frequency
of the $E_{g}$ phonon, the Drude peak is further renormalized by the EPC and
its spectral weight is reduced by a factor of (1+$\lambda ).$ Further
discussion can be found in Ref.\cite{Kuzmenko}.

\section{Conclusion}

MgB$_{2}$ is an unusual superconductor. It is not as far from conventional
materials as high-$T_{c}$ cuprates, or triplet Sr$_{2}$RuO$_{4}$. The
pairing symmetry is $s$, the driving force is electron-phonon interaction.
However, several factors distinguish MgB$_{2}$ from such more usual
superconductors as Nb, Nb$_{3}$Si, or even (B,K)BiO$_{3}$, to name a few.
The differences mainly stem from the fact that the charge carriers in MgB$%
_{2}$ fall into two distinctive groups: $\pi $-electrons, similar to those
in graphites, and $\sigma $-electrons, which represent highly unusual case
of covalent bands crossing the Fermi level. Only the latter group
demonstrate an anomalously strong interaction, and only with two phonons
with sufficiently small wave vectors.

This leads to a complex of uncommon features in the band structure,
transport properties, and superconductivity. In particular, the
superconducting state is characterized by two distinctively different order
parameters. Special symmetry of electronic states strongly suppresses the pair
scattering by impurities from one band system to the other, thus making the
two-gap superconductivity surprisingly unsensitive to sample quality. MgB$%
_{2}$ appears to be fairly unique, and, from our point of view, it is not
very likely that this compound can be optimized by a chemical modification
to raise substantially its critical temperature, as opposed, for example, to
high-$T_c$ cuprates.

\section{Acknowledments}

In course of our work on MgB$_2$ we enjoyed exceedingly fruitful
collaborations with outstanding researchers all over the world, whose names
appear in our joint publications ( Refs.\cite%
{us,Liu,us2,REVMGB2,MAZDHVA,NMRantrop,NMReva,imp}). We particularly
appreciate especially close collaborations with K. D.  Belashchenko (VA) and J.
Kortus (IM). We also wish to acknowledge very enlightening discussions with
O.K. Andersen, A.A. Golubov, O.V. Dolgov and the experimental group at the
Ames Laboratory.

VA acknowledges support from Ames Laboratory,
which is operated for the U.S.Department of Energy by Iowa State University
under Contract No. W-7405-82. IM
acknowledges support from the Office of Naval Research, and from the Kavli
Institute for Theoretical Physics, where part of this work was carried out,
under the National Science Foundation Grant No. PHY99-07949.

\end{multicols}

\begin{references}
\bibitem{CA} M. L. Cohen and P. W. Anderson, "Comments on the Maximum
Superconducting Temperature", in {\it Superconductivity in d- and f-metals,}
ed. by D.H. Douglass (AIP, New York), p. 17 (1972).

\bibitem{A}J. Nagamatsu, N. Nakagawa, T.
Muranaka, Y. Zenitani, and J. Akimitsu, Nature {\bf 410}, 63, (2001).
\bibitem{us} J. Kortus, I. I. Mazin, K. D. Belashchenko, V. P. Antropov and
L. L.~Boyer, Phys. Rev. Lett. {\bf 86}, 4656 (2001).


\bibitem{An} J. An and W. E. Pickett, Phys. Rev. Lett. {\bf 86}, 4366 (2001)

\bibitem{hirsch} J. E. Hirsch, Phys. Lett. A 282 , 392-398 (2001)

\bibitem{rice} K. Voelker, V. I. Anisimov and T. M. Rice, cond-mat/0103082

\bibitem{baskaran} G. Baskaran, Phys. Rev. {\bf B 65} (21) 212505 (2002)

\bibitem{yamaji} K. Yamaji, J. Phys. Soc. Jap. {\bf 70}, 1476 (2001)

\bibitem{marsiglio} F. Marsiglio, Phys. Rev. Lett. {\bf 87}, 247001 (2001).

\bibitem{Liu} A.Y. Liu, I. I. Mazin and J. Kortus, Phys. Rev. Let., {\bf 87}%
, 087005 (2001)

\bibitem{MGBSTR} S. Lee, H. Mori, T. Masui, Yu. Eltsev, A. Yamamoto and
S.~Tajima, J. Phys. Soc. Jap. {\bf 70}, 2255 (2001).

\bibitem{Bur} J. K. Burdett and G. J. Miller, Chem. Mater. {\bf 2}, 12
(1989).

\bibitem{Tupitsyn} I. I. Tupitsyn, Sov. Phys. Solid State {\bf 18}, 1688
(1976).

\bibitem{ArmstrongMgB2} D. R. Armstrong and P. G. Perkins, J.C.S. Faraday
II, {\bf 75}, 12 (1979).

\bibitem{Freeman-review} A. J. Freeman, A. Continenza, M. Posternak and S.
Massidda, in Surface Properties of Layered Structures, ed. G. Benedek
(Kluwer, Netherlands, 1992).

\bibitem{Medv} A. I. Ivanovskii and N. I. Medvedeva, Russ. J. Inorg. Chem. 
{\bf 45}, 1234 (2000).

\bibitem{us2} K. D. Belashchenko, M. van Schilfgaarde and V. P. Antropov,
Phys. Rev. {\bf B64}, 092503 (2001)

\bibitem{REVMGB2} V. P. Antropov, K. D. Belashchenko, M. van Schilfgaarde,
S. N. Rashkeev, in Studies of High Temperature Superconductors, {\bf 38},
91-116 (2002)

\bibitem{Freeman} H. J. F. Jansen and A. J. Freeman, Phys. Rev. B {\bf 35},
8207 (1987).

\bibitem{Harrison} W. A. Harrison, Electronic Structure and the Properties
of Solids (San Francisco, 1980).

\bibitem{CDexp} E. Nishibori, M.~Takata, M.~Sakata, H.~Tanaka, T.~Muranaka
and J.~Akimitsu, J. Phys. Soc. Jap. {\bf 70}, 2252 (2001).

\bibitem{REVIEW} N. B. Brandt, S. M. Chudinov and Ya. G. Ponomarev, Modern
problems in Condensed Matter Sciences, Vol. 20.1 (North-Holland,
Netherlands, 1988).

\bibitem{ARPES} H. Uchiyama, K. M. Shen, S. Lee, A. Damascelli, D. H. Lu, D.
L. Feng, Z.-X. Shen and S. Tajima,. Phys. Rev. Lett. ${\bf 88}$, 157002
(2002).

\bibitem{MGSURF} V. D. P. Servedio, S.-L. Drechsler, T. Mishonov, Int. J. of
Mod. Phys. {\bf B16}, 1613 (2002).

\bibitem{DHVAexp} E. A. Yelland, J. R. Cooper, A. Carrington, N. E. Hussey,
P. J. Meeson, S. Lee, A. Yamamoto and S. Tajima, Phys. Rev. Lett. {\bf 88},
217002 (2002).

\bibitem{Harima} H. Harima, Physica {\bf C 378-381}, 18 (2002).

\bibitem{Rosner} H. Rosner, J. M. An, W. E. Pickett, S.-L. Drechsler, Phys.
Rev. {\bf B66}, 024521 (2002); S. Elgazzar, P. M. Oppeneer, S.-L. Drechler,
R. Hayn and H.Rosner, Solid State Comm., 121, 99 (2002)

\bibitem{MAZDHVA} I. I. Mazin and J. Kortus, Phys. Rev. {\bf B65}, 180510
(2002).

\bibitem{Kong} Y. Kong, O. V. Dolgov, O. Jepsen and O. K. Andersen, Phys.
Rev. {\bf B64}, 020501 (2001).

\bibitem{Choi} H. J. Choi, D. Roundy, H. Sun, M. L. Cohen, and S. G. Louie,
Phys. Rev. B {\bf 66}, 020513 (2002); Nature {\bf 418}, 758 (2002).

\bibitem{URBANO} R. R. Urbano, P. G. Pagliuso, C. Rettori, Y. Kopelevich, N.
O. Moreno, and J. L. Sarrao, Phys. Rev. Lett. {\bf 89}, 087602 (2002).

\bibitem{SIMON} F. Simon, A. Janossy, T. Feher, F. Muranyi, S. Garaj,
L.Forro, Phys. Rev. Lett. {\bf 87}, 047002 (2001).

\bibitem{NMRantrop} K. D. Belashchenko, V. P. Antropov and S. N. Rashkeev,
Phys. Rev B, 64, 132506 (2001).

\bibitem{NMReva} E. Pavarini and I. I. Mazin, Phys. Rev. B, 64, 140504
(2001).

\bibitem{EXP1} H. Kotegawa, K. Ishida, Y. Kitaoka, T. Muranaka, J.~Akimitsu,
Phys. Rev. Lett. {\bf 87}, 127001 (2002).


\bibitem{EXP2} A. Gerashenko, K. Mikhalev, S. Verkhovskii, T. D'yachkova,
A.~Tyutyunnik, V. Zubkov, Appl. Magn. Reson. {\bf 21}, 157 (2001).

\bibitem{EXP3} J. K. Jung, S. H. Baek, F. Borsa, S. L. Bud'ko, G.~Lapertot,
P. C.~Canfield, Phys. Rev. B, 64(1), 012514 (2002).

\bibitem{EXP4} G. Papavassiliou, M. Pissas, M. Karayanni, M. Fardis, S.
Koutandos, and K. Prassides, cond-mat/0204238 (2002).

\bibitem{Budko} S. Bud'ko, G. Lapertot, C. Petrovic, C. E. Cunningham, N.
Anderson, and P. C. Canfield{\it .}, Phys. Rev. Lett. {\bf 86}, 1877 (2001)

\bibitem{Mgiso} D. G. Hinks, H. Claus, J. D. Jorgensen, Nature, 411, 6836
(2001)



\bibitem{Bohnen} K.-P. Bohnen, R. Heid and B. Renker, Phys. Rev. Lett. {\bf %
86}, 5771 (2001).

\bibitem{lastheat} Y. Wang, F. Bouquet, I. Sheikin, P. Toulemonde, B. Revaz,
M. Eisterer, H. W. Weber, J. Hinderer, A. Junod, cond-mat/0208169 (2002)

\bibitem{HEATEXP} F. Bouquet, Y. Wang, I. Sheikin, T. Plackowski, A. Junod,
S. Lee, S. Tajima, cond-mat/0207141 (2002).

\bibitem{MACHIDA} Y. Machida, S. Sasaki, H. Fujii, M. Furuyama, I. Kakeya
and K. Kadowaki, cond-mat/0207658 (2002).

\bibitem{oleg} In principle, near such a singularity the Migdal theorem is
violated and one has to solve the Dyson equation for the phonons
self-consistently (O.V. Dolgov, unpublished). However, since these
singularities are integrable it is not important for us in the moment.



\bibitem{Yildirim} T. Yildirim, O. Gulseren, J. W. Lynn, C. M. Brown, T. J.
Udovic, Q. Huang, N. Rogado, K. A. Regan, M. A. Hayward, J. S. Slusky, T.
He, M. K. Haas, P. Khalifah, K. Inumaru, and R. J. Cava, Phys. Rev. Lett. 
{\bf 87}, 037001 (2001).

\bibitem{Rodriguez} O. Rodriguez, A. I. Liechtenstein, I. I. Mazin, O.
Jepsen, O. K. Andersen, M. Methfessel, Phys. Rev.{\bf \ B42}, 2692 (1990).

\bibitem{0.51} The value reported for $\lambda _{ZZ}$ in Ref. \cite{Liu} was 
$\approx 0.6;$ it was later refined to be 0.51 (A.Y. Liu, private
communication).

\bibitem{Hui} J. C. Hui and P. B. Allen, J. Phys. {\bf F4}, L42 (1974).

\bibitem{Heid} R. Heid, Phys. Rev. {\bf B45}, 5052 (1992).

\bibitem{Crespi} V. H. Crespi and M. L. Cohen, Phys. Rev. {\bf B48}, 398
(1993).

\bibitem{shulga} S. V. Shulga, S.-L. Drechsler, H. Eshrig, H. Rosner and W.
E. Pickett, cond-mat/0103154.

\bibitem{junod} F. Bouquet, Y. Wang, R. A. Fisher, D. G. Hinks, J. D.
Jorgensen, A. Junod and N. E. Phillips, Europhys. Lett., {\bf 56}, 856
(2001).

\bibitem{Bascones} E. Bascones and F. Guinea, Phys. Rev. {\bf B64}, 214508
(2001)

\bibitem{Golheat} A. A. Golubov, J. Kortus, O. V. Dolgov, O. Jepsen, Y.
Kong, O. K. Andersen, B. J. Gibson, K. Ahn and R. K. Kremer, J. Phys.:
Condens. Matter {\bf 14}, 1353 (2002).

\bibitem{Suhl} H. Suhl, B. T. Matthias, and L. R.Walker, Phys. Rev. Lett. 
{\bf 3}, 552 (1959).

\bibitem{Moskal} V. A. Moskalenko, Fiz. Met. Met. {\bf 4}, 503 (1959).

\bibitem{Allen78} P. B. Allen and B. Mitrovic, in Solid State Phys., edited
by F. Seitz, D. Turnbull, and H. Ehrenreich ~Academic, New York, 1982, Vol.
37, p. 1.

\bibitem{golubov} A.A. Golubov and I.I. Mazin, Phys. Rev. B {\bf 55}, 15146
(1997).

\bibitem{imp} I. I. Mazin, O. K. Andersen, O. Jepsen, O. V. Dolgov, J.
Kortus, A. A. Golubov, A. B. Kuz'menko, D. van der Marel, Phys. Rev. Lett., 
{\bf 89}, 107002 (2002).

\bibitem{Junod-irr} Y. Wang, F. Bouquet, I. Sheikin, P. Toulemonde, B.
Revaz, M. Eisterer, H. W. Weber, J. Hinderer, A. Junod, cond-mat/0208169
(2002).

\bibitem{canf}R.A. Ribeiro, S.L.
Bud'ko, C. Petrovic, P.C. Canfield, cond-mat/0210530 (2002).
\bibitem{ARS} D. F. Agterberg, T. M. Rice, and M. Sigrist, Phys. Rev. Lett., 
{\bf 78}, 3374 (1997).

\bibitem{comment2} I. I. Mazin, O. K. Andersen, O. Jepsen, A. A. Golubov, O.
V. Dolgov, and J. Kortus, to be published.

\bibitem{schrieffer} J. R. Schrieffer. Theory of superconductivity. (1964),
300 pp.

\bibitem{PBA} F. J. Pinski, P. B. Allen, and W. H. Butler, Phys. Rev. {\bf %
B23}, 5080 (1981).02).

\bibitem{Kuzmenko} A. B. Kuzmenko, F. P. Mena, H. J. A. Molegraaf, D. van
der Marel, B. Gorshunov, M. Dressel, I. I. Mazin, J. Kortus, O. V. Dolgov,
T. Muranaka, J. Akimitsu, Solid State Comm. 121, 479-484 (2002)

\bibitem{comment} E. G. Maksimov, J. Kortus, O. V. Dolgov, and I. I. Mazin,
Phys. Rev. Lett. {\bf 89}, 129703 (2002).
\end{references}
\end{document}